\documentclass[aps,showpacs,superscriptaddress,preprint]{revtex4}%
\usepackage{amsfonts}
\usepackage{amsmath}
\usepackage{amssymb}
\usepackage{graphicx}%
\providecommand{\U}[1]{\protect\rule{.1in}{.1in}}

\begin{document}
\title{Ideal strengths and bonding properties of PuO$_{2}$ under tension}
\author{Bao-Tian Wang}
\affiliation{Institute of Theoretical Physics and Department of
Physics, Shanxi University, Taiyuan 030006, People's Republic of
China} \affiliation{LCP, Institute of Applied Physics and
Computational Mathematics, Beijing 100088, People's Republic of
China}
\author{Ping Zhang}
\thanks{Author to whom correspondence should be addressed. E-mail: zhang\_ping@iapcm.ac.cn}
\affiliation{LCP, Institute of Applied Physics and Computational
Mathematics, Beijing 100088, People's Republic of China}
\affiliation{Center for Applied Physics and Technology, Peking
University, Beijing 100871, People's Republic of China}

\begin{abstract}
We perform a first-principles computational tensile test on
PuO$_{2}$ based on density-functional theory within local density
approximation (LDA)+\emph{U} formalism to investigate its
structural, mechanical, magnetic, and intrinsic bonding properties
in the four representative directions: [001], [100], [110], and
[111]. The stress-strain relations show that the ideal tensile
strengths in the four directions are 81.2, 80.5, 28.3, and 16.8 GPa
at strains of 0.36, 0.36, 0.22, and 0.18, respectively. The [001]
and [100] directions are prominently stronger than other two
directions since that more Pu$-$O bonds participate in the pulling
process. Through charge and density of states analysis along the
[001] direction, we find that the strong mixed ionic/covalent
character of Pu$-$O bond is weakened by tensile strain and PuO$_{2}$
will exhibit an insulator-to-metal transition after tensile stress
exceeds about 79 GPa.
\end{abstract}
\maketitle

\section{INTRODUCTION}
Plutonium-based materials have been extensively studied due to their
interesting physical behaviors of the 5\emph{f} states and have
always attracted particular attention because of their importance in
nuclear fuel cycle \cite{Heathman,Atta-Fynn,Prodan2,Moore}.
Recently, experimental reports \cite{Haschke,Butterfield,Gouder} on
the strategies of storage of Pu-based waste illustrated that
metallic plutonium surface easily oxidizes to Pu$_{2}$O$_{3}$ and
PuO$_{2}$ in surrounding of air and moisture. Previous calculations
\cite{Prodan2} using the hybrid density functional of (Heyd,
Scuseria, and Enzerhof) HSE have established trends of the
electronic properties of actinide dioxides AnO$_{2}$ (An=Th$-$Es).
Along the series, Mott insulators ($f$$\rightarrow$$f$) are obtained
prior to PuO$_{2}$, whereas above AmO$_{2}$, charge-transfer
insulators (O 2$p\rightarrow$An 5$f$) are observed. PuO$_{2}$ and
AmO$_{2}$ lie in a crossover from localized to delocalized \emph{f}
electron character with increasing \emph{Z}. They have strong
5$f$-2$p$ orbital energy degeneracy, which can lead to unexpected
orbital mixing. Recent theoretical work performed by Petit \emph{et
al} \cite{Petit} demonstrated that the dioxide is the most stable
oxide for the actinides from Np onward. All these works indicate
that investigations of PuO$_{2}$ have particular meaning in actinide
compounds.

At ambient condition, PuO$_{2}$ crystallizes in a fluorite structure
with space group \emph{Fm$\bar{3}$m}. And at 39 GPa, PuO$_{2}$
undergoes a phase transition to an orthorhombic structure of
cotunnite type with space group \emph{Pnma} \cite{Dancausse}. In our
previous work \cite{SunJCP, Zhang2010, Shi2010}, we have
systematically investigated the structural, electronic, mechanical,
and optical properties of fluorite PuO$_{2}$ within the LDA+\emph{U}
and GGA+\emph{U} formalisms. We found that comparing with the
experimental data and the theoretical results, the accuracy of our
atomic-structure prediction for antiferromagnetic (AFM) PuO$_{2}$ is
quite satisfactory by tuning the effective Hubbard parameter
\emph{U} in a range of 3-4 eV within the LDA/GGA+\emph{U}
approaches. Subsequently, Jomard \emph{et al} \cite{Jomard}
confirmed our results and further reported optical and thermodynamic
properties of PuO$_{2}$. In 2008, Yin and Savrasov \cite{Yin}
successfully obtained the phonon dispersions of both UO$_{2}$ and
PuO$_{2}$ by employing the LDA+DMFT scheme. In 2009, Minamoto
\emph{et al.} \cite{Minamoto} investigated the thermodynamic
properties of PuO$_{2}$ based on their calculated phonon dispersion
within the pure GGA scheme.

However, although many efforts have been performed on PuO$_{2}$,
little is known on its theoretical tensile strength
\cite{Zhang2010}. The ideal strength of materials is the stress that
is required to force deformation or fracture at the elastic
instability.  Although the strength of a real crystal can be changed
by the existing cracks, dislocations, grain boundaries, and other
microstructural features, its theoretical value can never be raised,
i.e., the theoretical strength sets an upper bound on the attainable
stress. In this study, we focus our sight on the structural,
electronic, mechanical, and magnetic features of PuO$_{2}$ under
tension. The stress-strain relationships and the ideal tensile
strengths are obtained by performing a first-principles
computational tensile test (FPCTT) \cite{Zhang}. Under tension, the
bonding nature and electronic occupation characters are
systematically studied.

\section{computational methods}

The first-principles density-functional theory (DFT) calculations on
the basis of the frozen-core projected augmented wave method of
Bl\"{o}chl \cite{PAW} are performed within the Vienna \textit{ab
initio} simulation package (VASP) \cite{Kresse3}, where the LDA
\cite{LDA} is employed to describe electron exchange and
correlation. For the plane-wave set, a cutoff energy of 500 eV is
used. The \emph{k}-point mesh in the full wedge of the Brillouin
zone (BZ) is sampled by 9$\times $9$\times$9 grids according to the
Monkhorst-Pack \cite{Monk} for the fluorite unit cell and all atoms
are fully relaxed until the Hellmann-Feynman forces become less than
0.001 eV/\AA. The plutonium $6s^{2}7s^{2}6p^{6}6d^{2}5f^{4}$ and the
oxygen 2\emph{s}$^{2}$2\emph{p}$^{4}$ electrons are treated as
valence electrons. The strong on-site Coulomb repulsion among the
localized Pu 5\emph{f} electrons is described by using the
LDA+\emph{U} formalism formulated by Dudarev \emph{et al.}
\cite{Dudarev}. In this paper the Coulomb energy ($U$) and the
exchange energy ($J$) are set to be constants: $U$=4.75 eV and
$J$=0.75 eV. These values of $U$ and $J$ are the same as those in
our previous study of plutonium oxides \cite{SunJCP,Zhang2010}.
Using these parameters, the LDA+\emph{U} gives $a_{0}$=5.362 \AA ,
which is very close to the experimental value of 5.398 \AA\
\cite{Haschke}. And our results reproduce all the features included
in our previous work \cite{SunJCP}. In particular, we recover the
main conclusion that although the pure LDA fail to depict the
electronic structure, especially the insulating nature and the
occupied-state character of PuO$_{2}$, by tuning the effective
Hubbard parameter in a reasonable range, the LDA+\emph{U} approaches
can prominently improve upon the pure LDA calculations and, thus,
can provide a satisfactory qualitative electronic structure
description comparable with the photoemission experiments
\cite{Butterfield,Gouder}. Spin-polarized calculations are performed
and we find that the AFM spin alignment is the most stable
configuration among nonmagnetic, ferromagnetic (FM), and AFM
configurations. The total-energy difference
($E_{\text{FM}}\mathtt{-}E_{\text{AFM}}$ per formula unit at
respective optimum geometries) within the LDA+\emph{U} is calculated
to be 0.705 eV.

\begin{figure}[ptb]
\begin{center}
\includegraphics[width=0.8\linewidth]{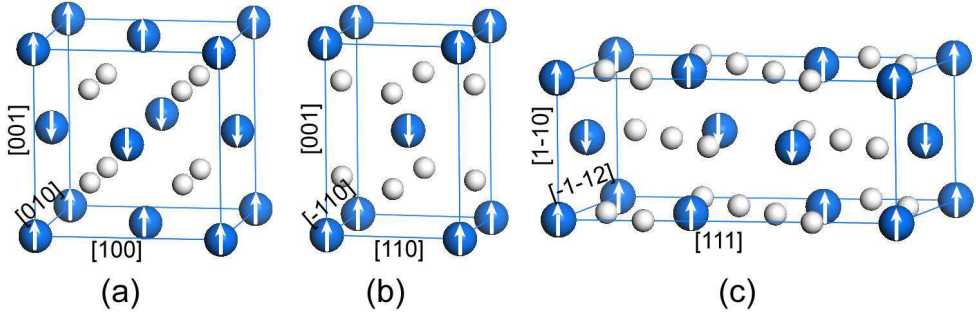}
\end{center}
\caption{(Color online) Schematic illustration of tension along (a)
[001] or [100], (b) [110], and (c) [111] orientations. Blue atoms
are plutonium atoms while white atoms are oxygen atoms. The AFM order is indicated by white arrows.}%
\label{tensile1}%
\end{figure}

In the FPCTT, the stress-strain relationship and the ideal tensile
strength are calculated by deforming PuO$_{2}$ crystal to failure.
The anisotropy of the tensile strength is tested by pulling the
initial fluorite structure along the [001], [100], [110], and [111]
directions. As shown in Fig. \ref{tensile1}, three geometric
structures are constructed to investigate the tensile strengths in
the four typical crystallographic directions: \ref{tensile1}(a)
shows a general fluorite structure of PuO$_{2}$ with four Pu and
eight O atoms; \ref{tensile1}(b) a body-centered tetrahedral (bct)
unitcell with two Pu and four O; and \ref{tensile1}(c) a
orthorhombic unitcell with six Pu and twelve O. In FPCTT, the
tensile stress is calculated according to the Nielsen-Martin scheme
\cite{Nielsen}
\begin{equation}
\sigma_{\alpha\beta}=\frac{1}{\Omega}\frac{\partial\\E_{\rm{total}}}{\partial\\\varepsilon_{\alpha\beta}},
\end{equation}
where $\varepsilon_{\alpha\beta}$ is the strain tensor ($\alpha,
\beta$=1,2,3) and $\Omega$ is the volume at the given tensile
strain. Tensile processes along the [001], [100], [110], and [111]
directions are implemented by increasing the lattice constants of
these three orientations step by step. At each step, the structure
is fully relaxed until all other five stress components vanish
except that in the tensile direction.

\section{results}

\subsubsection{Theoretical tensile strength}

\begin{figure}[ptb]
\begin{center}
\includegraphics[width=1.0\linewidth]{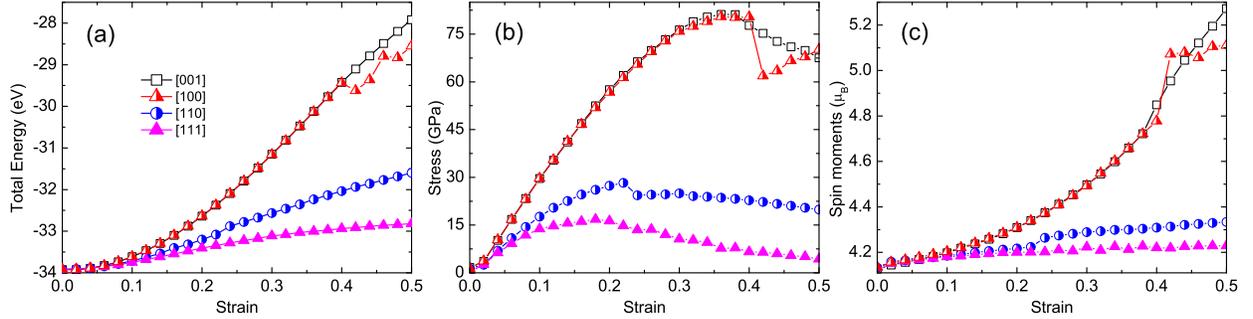}
\end{center}
\caption{(Color online) Dependence of the (a) total energy (per
formula unit), (b) stress, and (c) spin moments on tensile strain
for AFM PuO$_{2}$ in the [001], [100], [110], and [111] directions.}%
\label{tensile2}%
\end{figure}

\begin{figure}[ptb]
\begin{center}
\includegraphics[width=0.6\linewidth]{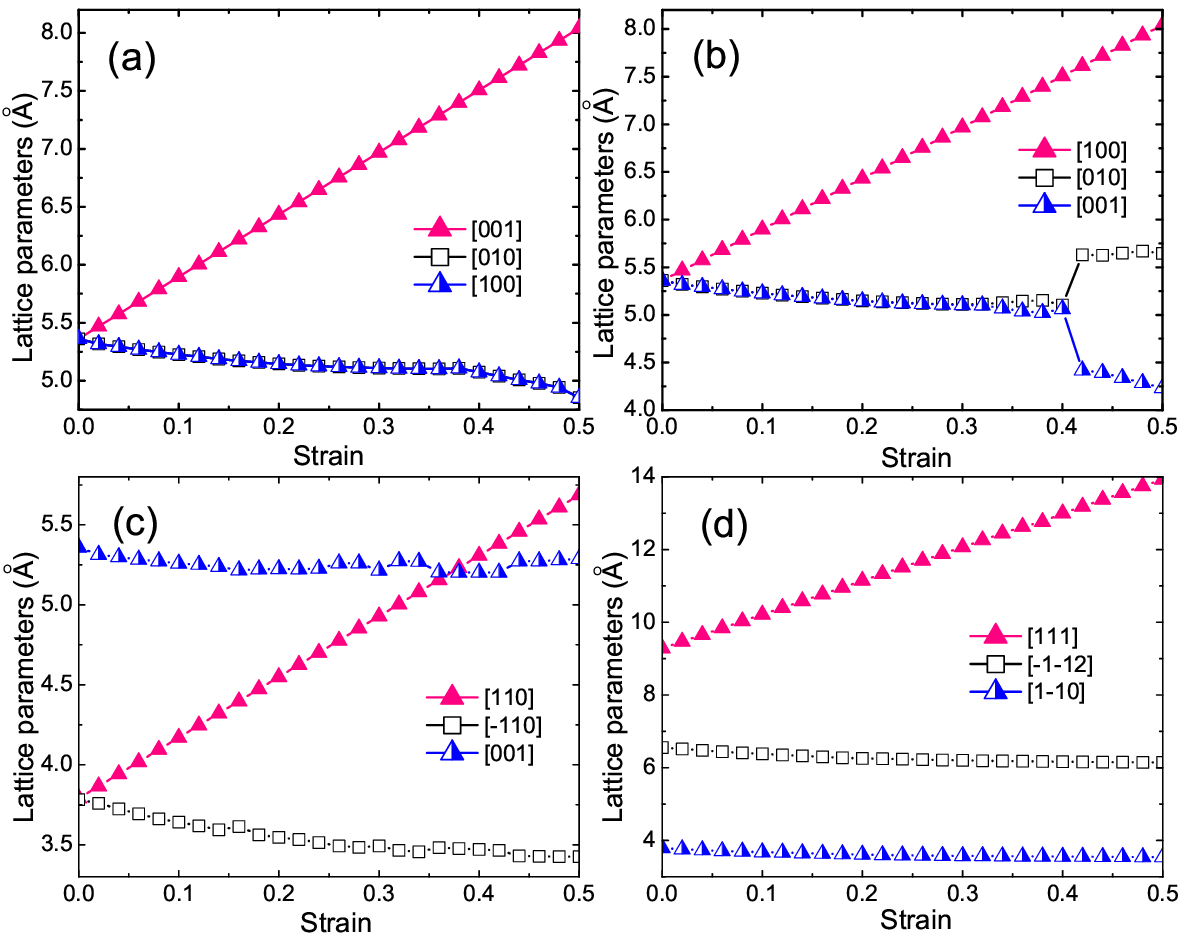}
\end{center}
\caption{(Color online) Dependence of the lattice parameters on
tensile strain for AFM PuO$_{2}$ in the (a) [001], (b) [100], (c) [110], and (d) [111] directions.}%
\label{tensile3}%
\end{figure}

The calculated total energy, stress, and spin moments as functions
of uniaxial tensile strain for AFM PuO$_{2}$ in the [001], [100],
[110], and [111] directions are shown in Fig. \ref{tensile2}.
Evolutions of the lattice parameters with strain in all four tensile
processes are presented in Fig. \ref{tensile3}. Clearly, all four
energy-strain curves increase with increasing tensile strain, but
one can easily find the inflexions by performing differentiations.
Actually, at strains of 0.36, 0.36, 0.22, and 0.18, the stresses
reach maxima of 81.2, 80.5, 28.3, and 16.8 GPa for pulling in the
[001], [100], [110], and [111] directions, respectively. These
results clearly indicate that the [001] direction is the strongest
tensile direction and [111] the weakest.

In fact, there are eight Pu$-$O bonds per formula unit for fluorite
PuO$_{2}$. The angle of all eight bonds with respect to the pulling
direction is 45$^{\circ}$ in [001] direction. For pulling direction
of [100], the bonding structure is same with that of [001]
direction. So their pressure behaviors of total energy, stress, and
spin moments are almost same before strain of 0.40. The special
behaviors of total energy, stress, and spin moments over
$\varepsilon$=0.40 for pulling in the [100] direction have tight
relations with the abrupt behavior of its crystal structure. In
fact, over $\varepsilon$=0.40 the face-centered tetragonal (fct)
structure will transit into an orthorhombic structure, as indicated
by Fig. \ref{tensile3}(b). The lattice parameter of [001] direction
is shortened and [010] direction elongated by tensile deformation.
However, in [110] direction only four bonds make an angle of
45$^{\circ}$ with the pulling direction. Four other bonds are
vertical to the pulling direction. In [111] direction, two bonds are
parallel to the pulling direction and six others make an angle of
about 19.5$^{\circ}$ with the pulling direction. It is easy to
understand that the bonds vertical to the pulling direction have no
contributions on the tensile strength and the bonds parallel to the
pulling direction are easy to fracture under tensile deformation.
Therefore, the fact that the tensile strength along the [001] or
[001] direction is stronger than that along [110] and [111]
directions is understandable. Besides, we note that the stress in
[110] direction experiences an abrupt decrease process after strain
up to 0.24. This is due to the fact that the corresponding four
Pu$-$O bonds (make an angle of 45$^{\circ}$ with the pulling
direction) have been pulled to fracture. The fracture behaviors have
been clarified by plotting valence electron charge density maps (not
shown). Under the same strain, the abrupt increase of spin moment
can be clearly seen [Fig. \ref{tensile2}(c)]. While the spin moments
in [110] and [111] directions only increase from 4.13 to 4.23 and
4.33 $\mu_{B}$, respectively, the spin moments in [001] and [100]
directions are increased up to about 5.27 and 5.11 $\mu_{B}$,
respectively, at the end of tensile deformation. In addition, the
evolutions of the lattice parameters with strain in Fig.
\ref{tensile3} clearly show that along with the increase of the
lattice parameter in the pulling direction, other two lattice
parameters vertical to the pulling direction are decreased smoothly
for tensile strains along [001], [110], and [111] directions. In
[001] direction, the evolutions of the lattice parameters along
[100] and [010] directions are absolutely same due to the structural
symmetry. For all these three tensile deformations, no structural
transition has been observed in our present FPCTT study. The
structural transition over $\varepsilon$=0.40 for [001] direction
tensile deformation is mainly due to the magnetic structure.

\subsubsection{Electronic properties under tension}

\begin{figure}[ptb]
\begin{center}
\includegraphics[width=0.8\linewidth]{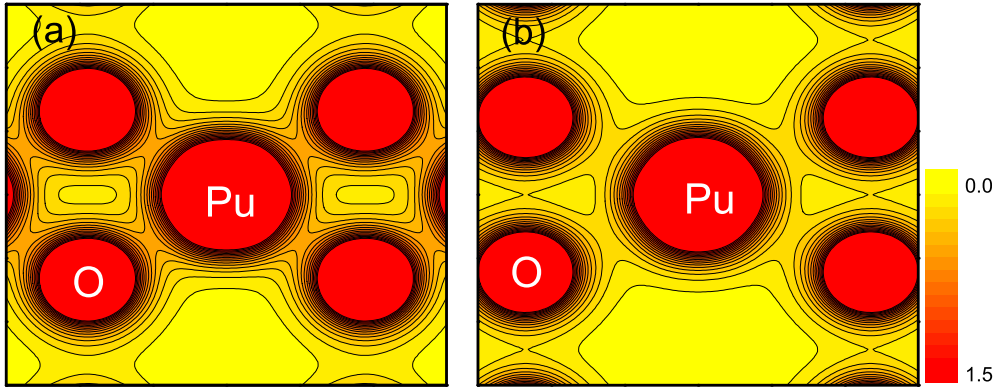}
\end{center}
\caption{(Color online) Valence charge density maps for AFM
PuO$_{2}$ in (01$\bar{1}$) plane under strains of (a) 0.0 and (b)
0.5 along [001] direction tensile strain. The contour lines
are drawn from 0.0 to 1.5 at 0.1 e/{\AA }$^{3}$ intervals.}%
\label{charge1}%
\end{figure}

\begin{figure}[ptb]
\begin{center}
\includegraphics[width=0.4\linewidth]{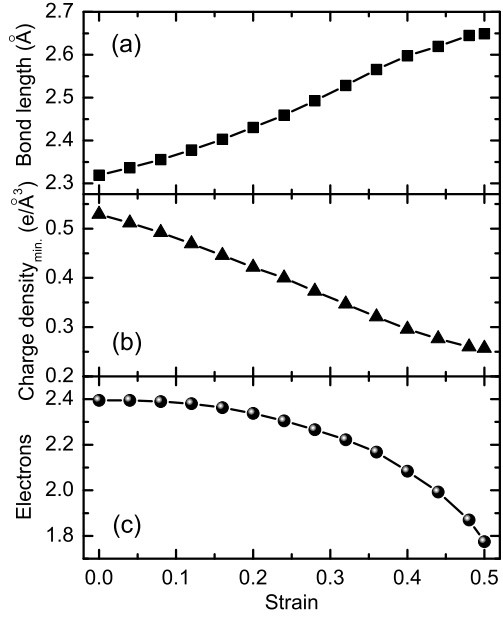}
\end{center}
\caption{(Color online) Dependence of (a) bond length of Pu$-$O
bonds in (01$\bar{1}$) plane, (b) correlated minimum values of
charge density along the bonds, and (c)number of electrons transfer
from each Pu to O atom
 on the tensile strain along the [001] direction.}%
\label{charge2}%
\end{figure}

In order to investigate carefully the physical nature of PuO$_{2}$
under tension, in the following we systematically study the
electronic structures of its AFM phase within LDA+\emph{U} formalism
for pulling along the [001] direction. The valence charge density
maps under strains of 0.00 and 0.50 along [001] direction tensile
deformation are plotted in Fig. \ref{charge1}. Clearly, the
interatomic distances in the [011] direction are elongated and that
in the [100] direction are shortened under tension. Under strain of
0.50, fractures of the Pu$-$O bonds can be seen. These results are
understandable based on our foregoing statements for its pressure
behaviors of total energy, stress, and lattice parameters. For
explicitly indicating the ionic/covalent character of PuO$_{2}$
under tension, we further plot in Fig. \ref{charge2} the evolutions
of Pu$-$O bond length in (01$\bar{1}$) plane, correlated minimum
values of charge density along the Pu$-$O bond, and number of
electrons transfer from each Pu to O atom under tensile deformation
along the [001] direction. The electron transfer analysis is
performed according to the Bader analysis \cite{Bader,Tang} and
similar ionic/covalent character investigations for thorium dioxide
and hydrides have been conducted in our previous works
\cite{WangThO2,WangThH}. Obviously, as indicated by Fig.
\ref{charge2}(a), the Pu$-$O bond length in (01$\bar{1}$) plane is
elongated by tensile stress in [001] direction. The initial minimum
value of charge density (0.53 $e$/{\AA }$^{3}$) along the Pu$-$O
bond, comparable to that along the Np$-$O bonds included in our
previous study of NpO$_{2}$ \cite{WangNpO2}, is prominently larger
than that along the Th$-$O bonds (0.45 e/{\AA }$^{3}$) in ThO$_{2}$
\cite{WangThO2}. This indicates that the Pu$-$O and Np$-$O bonds
have stronger covalency than the Th$-$O bonds. Under tension, this
minimum value of charge density for Pu$-$O bond decreases near
linearly to 0.26 e/{\AA }$^{3}$ at the end of tensile deformation,
which explicitly illustrates a decreasing behavior of covalent
feature for Pu$-$O bond. On the other hand, the number of electrons
transfer from each Pu to O atom under tensile deformation also
decrease with strain. Electrons of each atom are localized to their
dominated region and thus reduces the ionicity for PuO$_{2}$.
Overall speaking, the strong mixed ionic/covalent character of
Pu$-$O bond is weakened by tensile strain.

\begin{figure}[ptb]
\begin{center}
\includegraphics[width=0.4\linewidth]{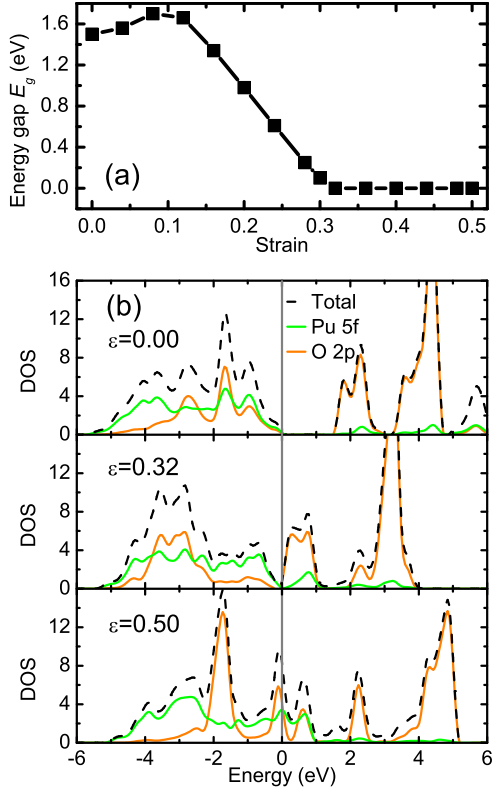}
\end{center}
\caption{(a) Dependence of the insulating band gap on the tensile
strain along the [001] direction. (b) Total DOS and PDOS under strains up to 0.00, 0.32, and 0.50. The Fermi energy level is set at zero.}%
\label{dos}%
\end{figure}

Figure \ref{dos}(a) shows the dependence of the insulating band gap
$E_{g}$ on the tensile strain along the [001] direction and
\ref{dos}(b) presents the total DOS as well as the projected DOS for
the Pu 5$f$ and O 2$p$ orbitals under strains up to 0.00, 0.32, and
0.50. Interestingly, the band gap varies smoothly under strain from
0.00 to 0.12 and then becomes to decrease near linearly from 1.66 eV
to zero under strain of 0.12 to 0.32. This explicitly indicates that
along the [001] direction tensile deformation PuO$_{2}$ will occur
an insulator-to-metal transition after tensile stress exceeds about
79 GPa. Figure \ref{dos}(b) further illustrates this conclusion. The
DOS at strain of 0.50 shows that the electrons of Pu and O atoms are
separated from hybridization feature to exhibit one-atom character.
And the main contribution for the metallic property is from both Pu
5$f$ and O 2$p$ orbitals.

\section{CONCLUSION}
In conclusion, we have studied the structural, mechanical, magnetic,
and electronic properties of PuO$_{2}$ under tension by
first-principles DFT within the LDA+\emph{U} formalism. Our
calculated ideal tensile strengths are 81.2, 80.5, 28.3, and 16.8
GPa, respectively, for pulling in the [001], [100], [110], and [111]
directions. The [001] and [100] directions are prominently stronger
than other two directions since that more Pu$-$O bonds participate
in the pulling process. While the spin moments in [110] and [111]
directions only increase a little, the spin moments in [001] and
[100] directions increase up to about 5.3 and 5.1 $\mu_{B}$,
respectively. Through charge and density of states analysis along
the [001] direction, we conclude that the initial strong mixed
ionic/covalent character of Pu$-$O bond is weakened by tensile
strain and PuO$_{2}$ will occur an insulator-to-metal transition
after tensile stress exceeds about 79 GPa. The main contribution for
the metallic property is from both Pu 5$f$ and O 2$p$ orbitals at
high strain domain.

\begin{acknowledgments}
This work was supported by NSFC under Grant No. 51071032, by the
National Basic Security Research Program of China, and by the
Foundations for Development of Science and Technology of China
Academy of Engineering Physics under Grant No. 2009B0301037.
\end{acknowledgments}

\end{document}